\documentstyle[12pt, psfig, natbib209]{article} 
\def\reference{\parskip 0pt\par\noindent\hangindent 0.5 truecm}

\def\arcdeg{\hbox{$^\circ$}}
\def\arcmin{\hbox{$^\prime$}}
\def\arcsec{\hbox{$^{\prime\prime}$}}
\def\ergcms{erg cm$^{-2}$ s$^{-1}$}
\begin{document}
%
%
\title{PFP~1: A Large Annular Planetary Nebula Caught in the First Stages of ISM Interaction}
%


\author{Mark J. Pierce,$^{1,2}$ 
 David J. Frew$^{2}$ and
 Quentin A. Parker$^{2,3}$ 
} 

\date{}
\maketitle

{\center
$^1$ Department of Physics, University of Bristol, Bristol, BS8 1TL, UK\\mark.pierce@bristol.ac.uk\\[3mm]
$^2$ Department of Physics, Macquarie University, Sydney, NSW 2109, Australia\\dfrew@ics.mq.edu.au\\[3mm]
$^3$ Anglo-Australian Observatory, PO Box 296, Epping, NSW 1710, Australia\\qap@ics.mq.edu.au\\[3mm]
}

%
\begin{abstract}

This paper presents (H$\alpha$ + [NII]) imaging and spectroscopy of a
previously unknown, highly evolved planetary nebula of low
excitation. It was discovered serendipitously from AAO/UKST H$\alpha$
Survey images as part of a project to exploit the survey data and has
evaded detection by previous surveys due to very low surface
brightness. It is a remarkably annular planetary nebula, some
19\arcmin\ across, making it one of the largest examples of its
type. We estimate a radius of 1.5 pc and a distance of 550pc as
derived from a new H$\alpha$ surface brightness -- radius
relation. PFP~1 has near perfect circular symmetry, broken only at the
NW edge which is coupled with significantly increased (H$\alpha$ +
[NII]) intensity. This indicates an early stage of an interaction with
the interstellar medium.  A blue central star candidate has been
identified from the SuperCOSMOS Sky Survey data.

\end{abstract}


\bigskip

%
%

\section{Introduction}

We report the discovery of a highly evolved, annular planetary nebula
(PN) of large angular size in the early stages of an interaction with
the interstellar medium (ISM) which we designate PFP~1. Old
interacting PNe are recognised as valuable probes of the ISM
(e.g. {Tweedy}, {Martos}, \& {Noriega-Crespo} 1995; {Xilouris}
{et~al.} 1996; {Kerber} {et~al.} 2000), but using them to build up a
global picture of the ISM properties is hampered by the lack of known
examples. This poor detection rate is due to the intrinsic low surface
brightness of these objects. The AAO/UKST H$\alpha$ Survey ({Parker}
\& {Phillipps} 1998, 2003) offers high quality narrowband images for
the whole of the Southern Galactic Plane (SGP) and therefore provides
a new tool which is uniquely suited to the detection of many more of
these large, diffuse PNe.

Because they are so difficult to detect, studies of these objects were
relatively neglected until recent years in favour of younger, brighter
PN. Early work by {Smith} (1976) looked at the effect of a PN-ISM
interaction on the position of the central star (CS) with respect to
the nebula. {Jacoby} (1981) was the first to present detailed
observational evidence of an interacting PN, Abell 35. However, it was
not until later, after analyses by {Borkowski}, {Sarazin}, \& {Soker}
(1990) and {Soker}, {Borkowski}, \& {Sarazin} (1991), that the
usefulness of interacting PNe as probes of the ISM itself was
recognised.

These initial studies sparked a renewed effort to find this kind of
object. {Tweedy} \& {Kwitter} (1996) present an imaging atlas of the
27 PNe of diameter $>$ 8\arcmin\ known at that time and find evidence
for PN-ISM interaction in all but four cases. More recent discoveries
have added to the haul (e.g. {Rauch} 1999; {Hewett} {et~al.} 2003),
taking the number of PNe known to be interacting with the ISM to over
100.

Most previous investigations to find large, faint PNe have relied on
careful scrutiny of broadband photographic survey images,
e.g. {Melmer} \& {Weinberger} (1990). While it has met with some
success, this method is not ideal because the narrow emission lines of
PN spectra are diluted in the broad bandpass of the survey
filters. This problem is particularly acute for PNe of really low
surface brightness and large angular extent. Use of narrowband imaging
around an emission line such as H$\alpha$ alleviates this problem, but
has until recently been limited to searches of relatively small areas;
e.g.  {Jacoby} \& {van de Steene} (1995) describe a search for new PNe
around PG1159 stars using narrowband imaging.

The AAO/UKST H$\alpha$ Survey is now available online ({Parker} \&
{Phillipps} 2003) and offers an unrivalled means of searching for this
kind of low surface brightness object. The survey comprises
contemporaneous narrowband H$\alpha$ and broadband R photographic
images which are scanned and digitised by the SuperCOSMOS plate
scanning machine at ROE ({Hambly} {et~al.} 2001), see
http://www-wfau.roe.ac.uk/sss/halpha/. This SuperCOSMOS H$\alpha$
Survey (SHS) is available both as 16$\times$ blocked down images and
full, 0.67\arcsec, resolution pixel data. The use of hypersensitised
Tech-Pan film ({Parker} \& {Malin} 1999) as the detector gives fine,
arcsec spatial resolution and extreme sensitivity to faint diffuse
emission (down to $\sim$5R \footnote{1 Rayleigh = 10$^{6}$/4$\pi$
photons cm$^{-2}$ s$^{-1}$ sr$^{-1}$ = 2.41 $\times$ 10$^{-7}$ erg
cm$^{-2}$ s$^{-1}$ sr$^{-1}$ at H$\alpha$}, Parker et al., in
preparation).

The survey data have already been successfully mined for small and
moderate-sized PNe via painstaking visual inspection of the survey
exposures to yield over 1000 new examples (e.g. {Parker} {et~al.} 
2003) in the Galaxy as a whole as well as over 300 new PNe in the
Galactic Bulge ({Peyaud}, {Parker}, \& {Acker} 2003). When completed,
this project will more than double the number known in the southern
hemisphere as compared with the combined work of the previous century.

The serendipitous discovery of PFP~1 from examination of blocked-down
H$\alpha$ data by MJP has powerfully demonstrated the utility of this
technique for discovering large ($>$ 4\arcmin) emission nebulosities,
and has led DJF and QAP to re-examine the entire H$\alpha$ survey in
this way. Rebinning the data to lower resolution
enhances the visibility of extended faint nebulosity. Of course, once
discovered, the full resolution pixel data can be examined which
permits finer study of the substructure of the PN and any interaction.
As a result more than 30 large PNe candidates were discovered (some up
to 30\arcmin\ diameter), as well as several new optical SNR candidates
(cf. {Parker}, {Frew}, \& {Stupar} 2004). Several of the PN
candidates have already been spectroscopically confirmed and will be
the subject of a separate paper (Frew \& Parker 2004, in preparation).

\section{Observations}

\subsection{AAO/UKST H$\alpha$ Survey Image}

\begin{figure}
\begin{centering}
\psfig{file=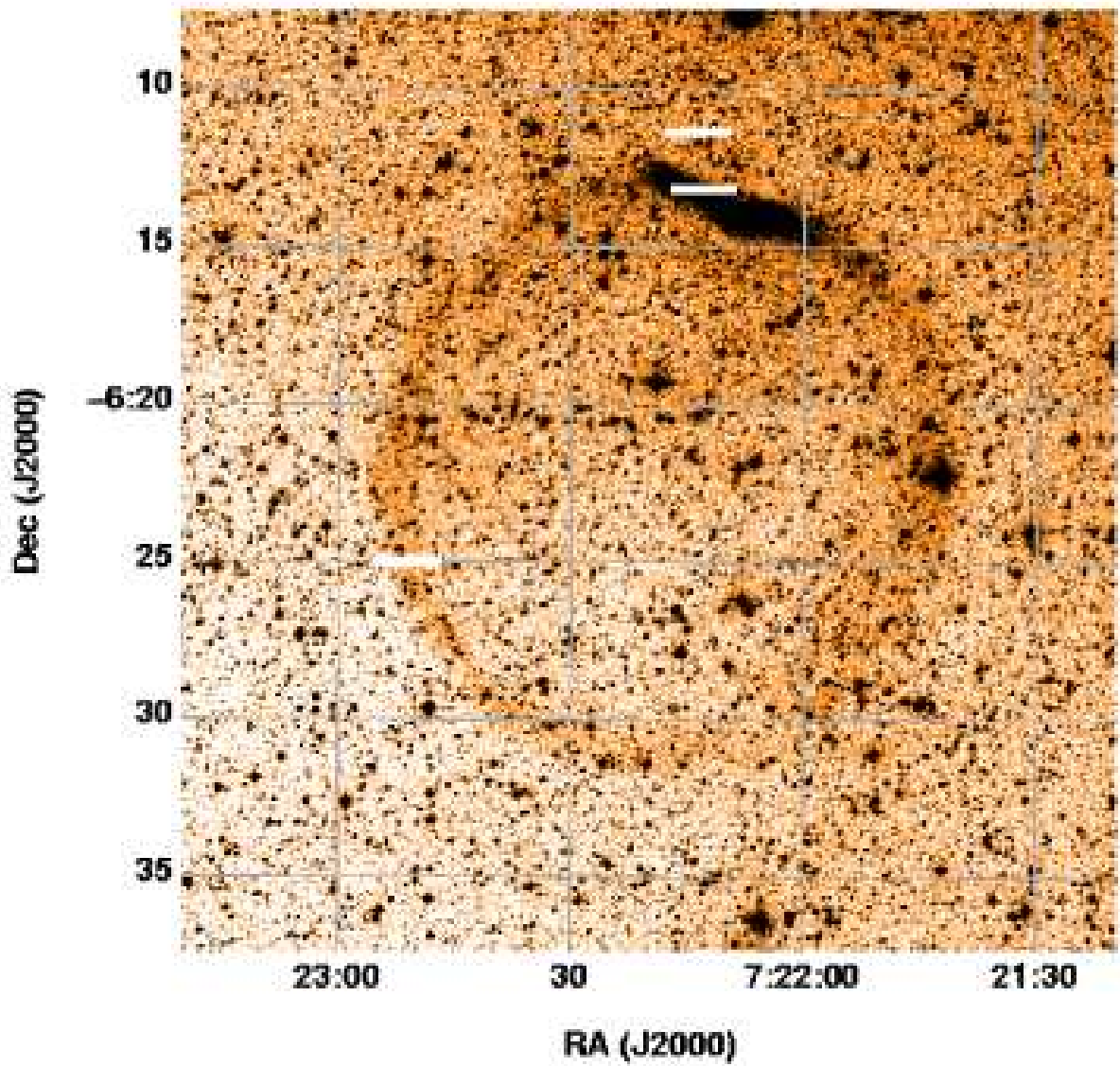,height=20cm}
\caption{Full resolution 30\arcmin\ by 30\arcmin\ SHS image of
PFP~1 (Quality reduced for preprint. Full resolution image will be
available in journal article). Several spectra of this object were
taken at the 1.9m telescope, SAAO, one of the faint east rim and three
of the brightened region as well as an offset sky spectrum. The slit
positions are indicated by the white lines.}
\end{centering}
\label{hanii}            
\end{figure}

PFP~1 was discovered in the NW corner of survey field HAL1109. We show
the full resolution 30\arcmin\ by 30\arcmin\ (H$\alpha$ + [NII]) SHS
image in Figure~\ref{hanii}. Most of the ring is extremely faint, with
the brightest part being the interacting northwest rim. The enhanced
limb itself can just be discerned on the POSS I and POSS II red plates
and the SERC ER survey as an elongated wisp of nebulosity (5\arcmin\
long) which is easily overlooked; there is little sign of the rest of
the 19\arcmin\ diameter annulus. The wisp is considerably fainter in
the blue, being barely visible at the plate limit on the SERC EJ
survey. This is a result of the relatively weak [OIII] emission from
this object as revealed by our spectra (Section 2.3).

From examination of various online multiwavelength data in the
vicinity of PFP~1, such as CO maps ({Dame} {et~al.} 1987), IRAS dust
maps ({Wheelock} {et~al.} 1994) and the SuperCOSMOS H$\alpha$ data
itself, it is clear that PFP~1 sits in a region of low extinction,
devoid of active star formation. The nearest large scale emission
within 50 square degrees is from the CMa OB1/R1 association which sits
6$^{\circ}$ south west at a distance of $\sim$1400pc ({Mel'Nik} \&
{Efremov} 1995). There is also a diffuse H$\alpha$ filament which
extends perpendicular to the plane at $l$ = 225$^{\circ}$ ({Haffner},
{Reynolds}, \& {Tufte} 1998) 3$^{\circ}$ south of the nebula, which
may link with general emission within the Galactic plane.

%
%
%
\begin{table*}
\centering
\caption{Basic data and derived quantities for PFP~1}
\label{pngeneral}
\begin{tabular}{@{}llllllll@{}}
\\
\hline
\\
\medskip
Name				&	PFP~1                          \\
\hline
Designation                    	&       PN G222.1+03.9                 \\
($\alpha, \delta$) J2000     	&       07 22 17.7, -06 21 46.4        \\
($l, b$)                	&       222.15, 3.94                   \\
SHS field               	&       HAL1109                        \\
Angular Dimensions            	&       1150\arcsec\ x 1100\arcsec\    \\ 
Excitation class        	&       2$^{a}$, 0.6$^{b}$            \\
Reddening, E(B-V)       	&       0.05$\pm$0.05                  \\
logF(H$\alpha$)         	&       --10.73$\pm$0.20   	       \\
logF(H$\beta$)          	&       --11.21$\pm$0.21   	       \\
Distance                	&       550 $\pm$220 pc \\
Nebular Diameter        	&       3.0 $\pm$1.2 pc                \\
Nebular Mass            	&       $\simeq$0.6 M$_{\odot}$	       \\
n$_{e}$(rms)			&	$\simeq$ 2 cm$^{-3}$          \\
Age                     	&       70 $\pm$30 x 10$^{3}$ yr       \\
\hline
\multicolumn{2}{l}{$^{a}$ Aller (1956)}\\
\multicolumn{2}{l}{$^{b}$ Dopita \& Meatheringham (1990)}\\
\end{tabular} 
\end{table*}

%

\subsection{SHASSA and VTSS Images}

The integrated H$\alpha$ flux for this object was determined from the
Southern H$\alpha$ Sky Survey Atlas data
(SHASSA, {Gaustad} {et~al.} 2001). SHASSA offers narrowband (H$\alpha$ +
[NII]) and continuum CCD images of the whole southern sky. The data
are at much lower spatial resolution (48\arcsec\ pixels) compared with
SHS (0.67\arcsec\ pixels) but have the benefit of being continuum
subtracted and intensity calibrated to a Rayleigh scale. Despite the
relatively coarse resolution of the SHASSA data, PFP~1 is such an
exceptionally large object that it is easily resolved by the survey.

These continuum subtracted data were used to measure the flux of both
the whole nebula as well as the bright interacting rim. The flux
measurement for the whole nebula was taken through a 20\arcmin\
circular aperture, while the bright region was fitted as tightly as
possible with an elliptical aperture. The Starlink routine {\sc
Aperphotom} was used in the {\sc Gaia} image analysis package to make
the measurements and the results are given in Table~\ref{pndata}. In
order to accurately account for the surrounding diffuse H$\alpha$
emission, measurements of sky background were made through an aperture
identical to the object aperture at fifteen points immediately
adjacent to the nebula. The scatter in these is our principal
uncertainty in the flux measurement.

The continuum subtracted SHASSA data come in two forms: raw,
48\arcsec\ resolution or data smoothed to 4\arcmin. The full
resolution SHASSA data shows some unphysical, negative pixel values
which appear to be residuals from poorly subtracted stellar images. To
check that these do not unduly influence the flux measurement, we
carried out the same procedure on SHASSA fields for 40 well-studied PN
(of a range of surface brightness) with an independent measure of
H$\alpha$ flux given in the literature. This confirmed that, despite
the artefacts, aperture photometry on the full resolution data returns
the best measurement of the H$\alpha$ flux which agrees with published
results to $\Delta$ F(H$\alpha$) = -0.001dex, $\sigma$ = 0.07 for
SHASSA minus literature fluxes.

The [NII]/H$\alpha$ ratios for the northern rim and the remainder of
the annulus (from the spectra described in Section 2.3) were used to
deconvolve the H$\alpha$ and [NII] contributions from the SHASSA red
flux, knowing the bandpass of the filter ({Gaustad} {et~al.} 
2001). Their quoted zero-point error of 9$\%$ and our estimated error
in the integrated [NII]/H$\alpha$ ratio for the whole nebula of
30$\%$, have been added quadratically to get the overall uncertainty
in the flux. We estimate logF(H$\alpha$)= --10.73$\pm$0.20 and use the
observed H$\alpha$/H$\beta$ ratio (from our spectra) to derive
logF(H$\beta$) = --11.21$\pm$0.21.

We also looked at the relevant field (Mon07) from the Virginia Tech
Spectral-Line Survey (VTSS; {Dennison}, {Simonetti}, \& {Topasna}
1998).  The VTSS resolution is coarser than SHASSA (1.6\arcmin\
pixels) and consequently it is more difficult to adequately subtract
the diffuse background.  On the other hand, the H$\alpha$ filter has a
narrower bandpass and the [NII] contribution can be ignored.  However,
{Gaustad} {et~al.} (2001) found the VTSS data to be fainter than
SHASSA by a factor of 1.25 in one field, while {Finkbeiner} (2003)
found this offset to be applicable to all fields in common.  We
confirm these findings by estimating H$\alpha$ fluxes from VTSS for 9
well-studied PNe on different fields, finding an offset of 0.09 dex
(VTSS fainter).  We applied this correction to the integrated flux for
PFP 1 estimated from the VTSS image and get logF(H$\alpha$) =
--10.8$\pm$0.3, where the error is dominated by the uncertain
background subtraction.  Nevertheless, this is excellent agreement
with our [NII]-corrected flux derived from SHASSA.

%
%
\begin{table*}
\centering
\caption{Log of spectroscopic observations.}
\label{log}
\begin{tabular}{@{}cr@{ }lr@{s}l}
\\
\hline
\\
\medskip
Date (UT) &Slit&position  &Expo&ure	        \\
\hline                                                      
13 02 2004&NW rim - 07$^{h}$&22$^{m}$ 13$^{s}$ -06\arcdeg\ 13\arcmin\ 05\arcsec\	&1200			\\
14 02 2004&NW rim - 07$^{h}$&22$^{m}$ 13$^{s}$ -06\arcdeg\ 13\arcmin\ 05\arcsec\	& 900& $\times$ 2	\\
14 02 2004& E rim - 07$^{h}$&22$^{m}$ 51$^{s}$ -06\arcdeg\ 25\arcmin\ 00\arcsec\	&1200&          	\\
14 02 2004&Offset Sky - 07$^{h}$&22$^{m}$ 14$^{s}$ -06\arcdeg\ 11\arcmin\ 20\arcsec\ & 900&			\\
\hline
\end{tabular}
\end{table*}

%

\subsection{Spectroscopic Observations}

%
%
\begin{table*}
\centering
\caption{Line fluxes and ratios for PFP~1 from combined spectrum of
interacting northwest rim.}
\label{pndata}
\begin{tabular}{@{}llllllll@{}}
\\
\hline
\\
\medskip
ID   &   Wavelength (\AA) & F($\lambda$) & I($\lambda$)              \\
\hline

$[$OII$]$ &3727     	&	  585 	&  611   	&       \\
$[$NeIII$]$&3869    	&	    * 	&    *    	& 	\\
H$\gamma$ &4340     	&	   45 	&   46   	&       \\
H$\beta$  &4861     	&	  100 	&  100   	&       \\
$[$OIII$]$&4959     	&	   50 	&   50   	&       \\
$[$OIII$]$&5007     	&	  141 	&  140   	&       \\
$[$OI$]$  &6300	    	&           * 	&    *    	&       \\
$[$OI$]$  &6364	    	&           * 	&    *    	&       \\
$[$NII$]$ &6548 $^{a}$	&	  228 	&  216   	&       \\
H$\alpha$ &6563 $^{a}$	&	  301 	&  285   	&       \\
$[$NII$]$ &6584 $^{a}$	&	  727 	&  688   	&       \\
$[$SII$]$ &6717 $^{a}$	&	  133 	&  125   	&       \\
$[$SII$]$ &6731 $^{a}$	&	   85 	&   80   	&       \\
$[$NII$]$/H$\alpha$&	&  		& 3.17$\pm$0.20 &	  \\
$[$OIII$]$/H$\beta$&	&  		& 1.90$\pm$0.10 &	  \\
$[$SII$]$ 6717/6731&	&  		& 1.57$\pm$0.05 &	  \\
\hline	
\multicolumn{4}{l}{F(H$\beta$) = 1.96 $\times 10^{-14}$  \ergcms, c(H$\beta$) = 0.07} \\
\multicolumn{4}{l}{$^{a}$ From gaussian deblends of close lines.}		\\
\multicolumn{4}{l}{* Present but faint or affected by sky lines.}		\\

\end{tabular}
\end{table*} 
%
%
%
\begin{figure}
\begin{centering}
\psfig{file=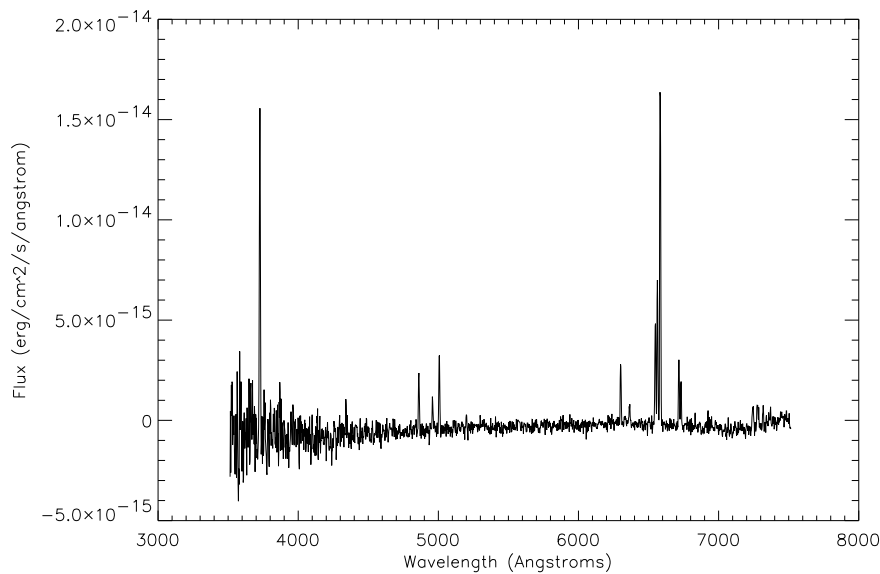,height=8cm}
\caption{The combined, flux calibrated spectrum of the NW rim.}
\label{spec}
\end{centering}
\end{figure}

%

Spectra were obtained with the SAAO 1.9m telescope on the nights of
Feb 13 and Feb 14, 2004. The standard grating spectrograph was used
with a SITe 1798 x 266 pixel CCD coupled to an f/2.2 camera with an 86
mm beam. Grating 7 was used (300 lines/mm), giving a spectral coverage
of 3600--7400\AA. On-chip binning (2x) perpendicular to the dispersion
direction was used to reduce readout noise. The slit width was widened
to 400$\mu$m ($\sim$2.4\arcsec\ on the sky) to allow greater
throughput for the faint emission of the extended PNe observed on this
run, though with some loss of resolution (to $\sim$7\AA). The
unvignetted slit length perpendicular to the dispersion was
1.6\arcmin\ and was fixed in an east-west direction. Spectra of both
the faint eastern edge of the PN shell and the bright rim were
obtained in separate observations summarised in Table~\ref{log}. An
offset sky frame was also taken as the nebula subtended the entire
slit. The positions of the slit are shown in Figure~\ref{hanii}. Dome
screen flats from a quartz-iodine lamp and twilight flats were also
obtained together with observations of spectrophotometric standard
stars ({Stone} \& {Baldwin} 1983) and PN flux standards from the list
of {Dopita} \& {Hua} (1997). Wavelength calibration was applied via
Cu-Ar arc exposures bracketing the observations. The spectra were
reduced and examined using standard IRAF routines.

The two spectra of the NW rim taken on Feb 14 were combined and used
for the analysis which follows. As the conditions on Feb 13 were not
photometric, the spectrum taken on that night was examined separately
and confirms the findings from the two combined spectra. The combined
spectra are plotted in Figure~\ref{spec}. 

Table~\ref{pndata} includes the measured flux values, normalised to
H$\beta$ = 100 in column 2 and the fluxes corrected for reddening in
column 3, (assuming H$\alpha$/H$\beta$ = 2.85; {Brocklehurst} 1971)
and adopting the R= 3.1 reddening law of {Savage} \& {Mathis}
(1979). The line intensities are estimated to be accurate to 30 $\%$,
and as an internal check, the ratios [OIII]
$\lambda$5007/$\lambda$4959 = 2.8 and [NII]
$\lambda$6584/$\lambda$6548 = 3.2, agree well with the theoretical
values of each. We derive the logarithmic extinction at H$\beta$, c =
0.07, which leads to E(B-V)= 0.05 and a total visual absorption,
A$_{V}$= 0.15.

\subsection{Identification of Candidate Central Star}

\begin{figure}
\begin{centering}
\psfig{file=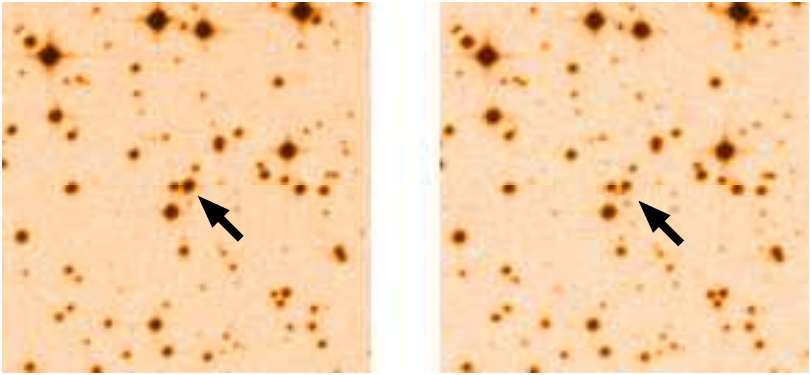,height=5cm}
\caption{Candidate central star at 07$^{h}$ 22$^{m}$ 17.7$^{s}$
-06\arcdeg\ 21\arcmin\ 46.4\arcsec\ in UKST B$_{J}$ (left) and R band
(right) images. Each image is 3\arcmin\ on a side. The candidate CS is
highlighted in each case by the arrow.}
\label{csims}
\end{centering}
\end{figure}

%
%
%
\begin{table*}[ht]
\centering

\caption{Magnitudes and colours for the candidate central star which
sits at the geometric centre of the nebula. B$_J$ magnitude is
estimated from the SuperCOSMOS Sky Survey data as described in Section
2.4 }

\label{csdata}
\begin{tabular}{@{}lllll@{}}
\\
\hline
\\
\medskip
Band	&	Source	&	magnitude	\\
\hline						  
B$_J$*&	SSS	&	15.43 $\pm$0.15	\\
R$_{2}$	&	SSS	&	15.45 $\pm$0.15	\\
I	&	SSS	&	16.51 $\pm$0.15	\\
J	&	2MASS	&	16.64 $\pm$0.16	\\
\hline
Colour	&		&			\\
\hline
B$_J$--R&		&	--0.02		\\
R--I	&		&	--1.06		\\
B$_J$--J&		&	--1.21		\\
\hline
\end{tabular}
\end{table*}

In order to find potential central star candidates, we examined
digitised UKST R and B$_J$ plates from the SuperCOSMOS Sky Survey
(hereafter SSS, {Hambly} {et~al.} 2001, see
http://www-wfau.roe.ac.uk/sss/index.html). A careful
visual inspection revealed an object just 15\arcsec\ from the
geometric centre of this large nebula which is brighter in the blue
plate than the red when compared with other stars in the field. The
two images of this object are shown in Figure~\ref{csims}, first in
B$_J$ then in R. North is up and east is left.

The CS should still be close to the centre of PFP~1; only the
northwest rim is brightened, implying that the interaction is at an
early stage and that the bulk of the nebula is still moving with the
CS. Objects showing more advanced interaction may have a substantially
displaced CS, e.g. Sh~2-174 ({Tweedy} \& {Napiwotzki} 1994) where the
CS has actually left the nebula, and Sh~2-216 ({Tweedy} {et~al.} 1995)
where the CS is displaced by 24\arcmin\ from the geometric centre of
this 1.6$^{\circ}$ PN, due to the nebular material being slowed and
ultimately reabsorbed into the ISM.

Some additional work was required to determine the B$_J$ magnitude
from the SSS data. The SuperCOSMOS detection and parameterisation
software had not deblended the candidate CS from its nearest neighbour
to the east in the blue data, although this pair were separated in
R. We used the Starlink routine {\sc SExtractor} in conjunction with
{\sc Gaia} to deblend these two objects from the UKST B$_J$ pixel
data. As {\sc SExtractor} provides a flux measurement from aperture
photometry for each object, we were able to make flux measurements for
$\sim$2000 stars within 10\arcmin\ of our candidate CS. This formed a
well constrained B$_J$ -- flux relation which could be used to
calibrate the B$_J$ magnitude of the now deblended CS candidate based
on its measured flux. Fitting to the calibration curve in this way
gave us a magnitude estimate of B$_{J}~\sim$15.43~$\pm$0.15 mag for
our favoured CSPN.  This magnitude is consistent with those tabulated
in the USNO-B1.0 and GSC 2.2.01 surveys.

In Table~\ref{csdata} we show photometric measurements of the
candidate CS obtained from the SSS, as well as the J (1.6$\mu$m)
magnitude from the Two Micron All Sky Survey (2MASS) Point Source
Catalog. Along with the blue B$_J$--R colour of $\sim$ 0.0, the
B$_J$--J colour of --1.2 indicates that this is a very blue
object, and is consistent with the predicted B--J~=~--1.1 for a
10$^{5}$K CSPN, giving us confidence that we have indeed identified
the true central star.

An apparent blue magnitude of $\sim$15.4 is quite a reasonable value
for an evolved CSPN at the distance and reddening that we derive for
PFP~1. The calculated M$_{B_{J}}$~$\simeq$~6.6~$^{+1.1}_{-0.9}$ is
similar to the central star of the similarly large PN Sh~2-216, which
has M$_{B}$~=~6.5 from an accurately determined trigonometric parallax
({Harris} {et~al.} 1997).

%

\section{Analysis}

\subsection{Distance and Mass}

In the absence of definitive photometric and spectroscopic information
on the candidate central star, the distance can be estimated by the
classical Shklovsky method, assuming a value for the ionised mass of
the shell. Inverting this procedure allows a determination of ionised
masses for a number of evolved PNe of similar evolutionary age and
morphology to our object, with distances taken from the
literature. Ionised masses were calculated using the formula of
Pottasch (1996):

\begin{equation}
\label{eq:pott}
M = 4.03 \times 10^{-4} \epsilon^{1 \over 2} d^{5 \over 2} (H\beta)^{1 \over 2} \Theta^{3 \over 2} M_{\odot}
\end{equation}

where $d$ is the distance in kpc, $H\beta$ is the flux in units of
10$^{-11}$ \ergcms, and $\Theta$ is the nebular radius in arcseconds.
We have assumed a filling factor, $\epsilon$ = 0.3 (cf. {Mallik} \&
{Peimbert} 1988; {Boffi} \& {Stanghellini} 1994; {Pottasch} 1996). The
range of ionised masses for 25 low surface brightness PNe with good
distance determinations is 0.06 to 1.3 solar masses, excluding the
anomalous objects PHL 932 ({Mendez} {et~al.} 1988; {Acker} {et~al.} 
1998) and K 1-27 ({Rauch}, {Koeppen}, \& {Werner} 1994; {Ciardullo}
{et~al.} 1999). This brackets the distance of PFP~1 to between 240 and
770 pc.

Since the more massive (evolved) PNe are known to have
fainter central stars, the difference between the central star
magnitude and the integrated H$\beta$ magnitude of the nebula is an
indicator of the ionised mass of the object, and functions as a
distance-independent parameter.  We have found for evolved PNe
 with logS($H\beta$) $\leq$ --3.5 \ergcms\ sr$^{-1}$ that $\Delta$ =
m$_{B,CS}$ -- m$_{H\beta,neb}$ ranges between --4.0 and +4.0 and
correlates quite well with ionised mass.  The H$\beta$ magnitudes were
derived using the formula of {Jacoby} (1989) and are corrected for
reddening.  Note that for objects that suffer moderate reddening (or
less), the apparent magnitudes may be used with little loss in
utility.  We have found $\Delta$ $\leq$--1.0 for low-mass,
high-excitation, optically-thin PNe and that $\Delta$ is generally
positive for the larger, more massive PNe with fainter, more rapidly
evolving central stars. The mass of our PN is probably not as low as
0.1 M$_{\odot}$, a value typical of high-excitation objects such as
NGC 246, NGC 7094 and Abell 36 ({Kaler} 1981), which have relatively
luminous central stars, and presumably lower progenitor star
masses. For PFP~1, we derive m$_{H\beta,neb}$ = +14.1 and $\Delta$ =
+1.2, suggesting our PN has a relatively faint central star (for its
observed H$\beta$ flux) and therefore a moderately high ionised mass
and large intrinsic diameter, consistent with the low excitation
class.  Hence the distance is unlikely to be near the lower bound of
our bracketed range.

\begin{figure}
\begin{centering}
\psfig{file=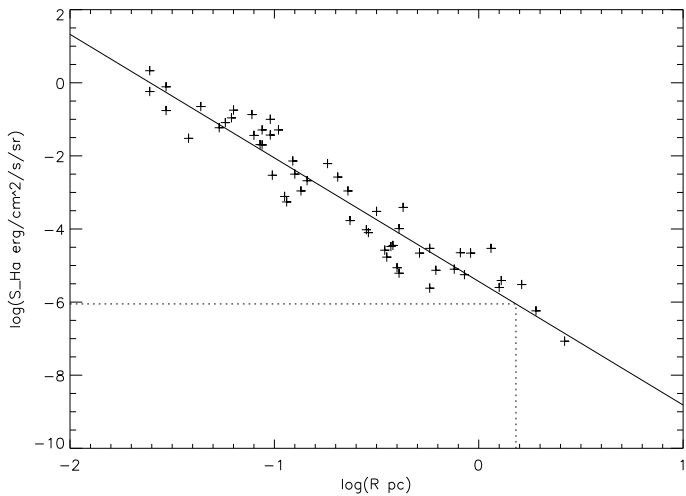,height=7cm}

\caption{An empirical dereddened H$\alpha$ surface brightness -- radius
relation for planetary nebulae based on 53 local calibrators with
well-determined distances selected from the literature. The
calibrating PNe range in brightness from young objects like NGC 7027
through to faint examples similar to our new PN. The best fit is
represented by the equation: log S(H$\alpha$) = --5.43 -- 3.38
$\times$ log(R) and is consistent with an R$^{-3}$ law. The position
of PFP~1 is indicated by the dashed lines.}

\label{pndist}
\end{centering}
\end{figure}

An alternative distance method is to use an empirical H$\alpha$
surface brightness -- radius relation, calibrated with PNe which have
accurate distance estimates determined using a primary
method. H$\alpha$ fluxes for these PNe were taken from the literature,
supplemented with flux measurements derived from SHASSA data (see
Section 2.2) for a number of low surface brightness objects. The log
of the radius (derived from the quoted distance) can be plotted
against the log of the reddening corrected surface brightness to give
a well defined trend which can be used as a distance indicator
accurate to ~40$\%$. A preliminary version of this plot is shown in
Figure~\ref{pndist} and is broadly consistent with an R$^{-3}$ law as
found for Magellanic Cloud PNe by {Stanghellini} {et~al.} (2003). Full
details of the calibration will be published elsewhere (Frew \&
Parker, in preparation).

From our estimated dereddened log S(H$\alpha$) = --6.05, we derive a
radius, r= 1.5 $\pm$0.6 pc, and the distance, D = 550 $\pm$220 pc,
immediately follows.  The estimated ionised mass is then 0.6
M$_{\odot}$ (for an assumed $\epsilon$ = 0.3) comfortably within the
range calculated for other highly-evolved PNe. We intend to improve
our distance estimate by obtaining accurate UBV photometry and a
spectrum of the CSPN candidate.

\subsection{Other Nebular Parameters}

We used the combined spectrum of the NW rim to estimate the principal
nebular parameters, tabulated in Table~\ref{pngeneral}.  We were also
able to measure an [NII]/H$\alpha$ ratio in the much fainter eastern
rim for comparison purposes (see Section 4). The derived reddening,
E(B--V) = 0.05, can be compared to the asymptotic extinction of
E(B-V)= 0.26 (A$_{V}$= 0.81) in this direction, as determined from the
exinction map of {Schlegel}, {Finkbeiner}, \& {Davis} (1998). This is
an unusually transparent sightline through the disk, despite the low
galactic latitude ($b$ = 3.9\arcdeg), so no accurate distance estimate
is possible using the extinction-distance method.

The surface brightness at H$\alpha$ derived from the brightest 71
binned pixels (99 arcsec) along the slit is log S(H$\alpha$)= --4.97,
which can be compared with logS(H$\alpha$)= --5.43 derived using the
brightest SHASSA pixel value, after deconvolving the [NII]
contribution to the SHASSA data and correcting for background.  There
is a 30$\%$ probable error on the H$\alpha$ flux through the slit
which sampled the brightest portion of the northern rim, which is only
about 35\arcsec\ north to south.  Considering that each 48\arcsec\
SHASSA pixel in the vicinity would sample nebulosity of varying
surface brightness, we consider these values to be in reasonable
agreement. The interacting rim is therefore a factor of 5--10 higher in
surface brightness than that averaged over the entire nebula.

The excitation class is low, only 2 on the scale of {Aller} (1956), as
revealed by the relatively strong [NII] and [OII] lines and weak
[OIII] emission seen in our spectra. The spectrum shows very strong
similarity to the bright eastern section of Sh2-216 ({Fesen}, {Blair},
\& {Gull} 1981), though the [NII] emission is somewhat stronger; this
may be an abundance effect. A decimal excitation class of 0.6 was
calculated according to {Dopita} \& {Meatheringham} (1990), Equation
2.1a:

\begin{equation}
\label{eq:ec}
E.C. = 0.45[F(5007)/F(H\beta)]
 \end{equation}

Both indices indicate a nebula of low excitation, as
expected for a highly evolved PN of large intrinsic diameter
surrounding a faint evolved central star at the beginning of the white
dwarf cooling track.

From the [SII]6717/6731 ratio of 1.57 we can only find the limit for
n$_{e} <$ 10$^{2}$ cm$^{-3}$ ({Stanghellini} \& {Kaler} 1989) because
the [SII] ratio is insensitive to electron density at this value.  We
can also calculate an approximate value of n$_{e}$ from the H$\alpha$
flux, radius and distance of the object:

\begin{equation}
\label{eq:n_e}
n_e^2 (rms) = 2.8 \times 10^{22} {F(H\alpha) \over \theta^3 D} ,
\end{equation}

where D is in pc and $\theta$ is the radius in arcsec.  We get
n$_{e}\sim$ 2cm$^{-3}$, which is one of the lowest values known.

We are unable to calculate the electron temperature due to the
non-detection of the [OIII]$\lambda$4363 and the [NII]$\lambda$5755
lines.

The kinematic age can be crudely estimated assuming an expansion
velocity of 20 kms$^{-1}$ ({Borkowski} {et~al.} 1990). The age is then
$\sim$70x10$^{3}$ yr although this is a lower limit because the
expansion rate is likely to have slowed over the course of its
evolution.

Although we only have low dispersion spectra, using the emission line
fitting routines in the RVSAO package in IRAF leads to a heliocentric
corrected radial velocity of --18 $\pm$19 kms$^{-1}$, based on fits to
8 lines. Comparing equivalent measures for two {Dopita} \& {Hua} (1997)
PNe standards observed on the same night, we find agreement to $\pm$
20 kms$^{-1}$.

\section{ISM Interaction}

\begin{figure}
\begin{centering}
\psfig{file=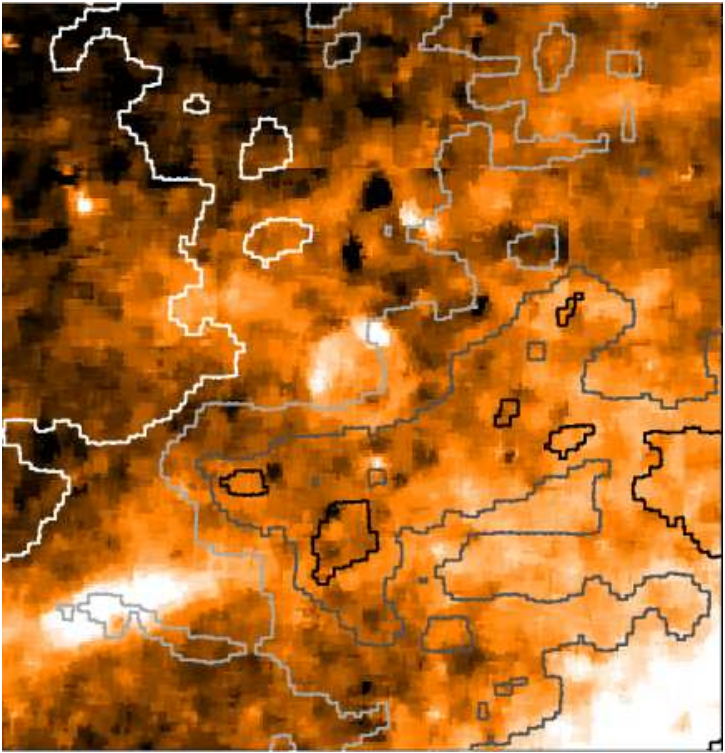,height=10cm}

\caption{IRAS 100$\mu$m contours overlaid on a $\sim$ 2$^{\circ}$ by
2$^{\circ}$ area of smoothed SHASSA H$\alpha$ data. The IRAS contours
are at 10 (white), 12, 14, 16 (black) MJy sr$^{-1}$} 

\label{shassairas}
\end{centering}
\end{figure}

{Tweedy} \& {Kwitter} (1996) suggest three criteria to confirm that PN
asymmetry is caused by a PN-ISM interaction. They propose that such
interactions are characterised by an asymmetry in the outer region of
the nebula, accompanied by an increase in brightness at the
interacting edge which also exhibits an increased [NII]/H$\alpha$
ratio with respect to the rest of the nebula. This increased
[NII]/H$\alpha$ ratio results from material being compressed at the
PN-ISM boundary, raising the density which, in turn, increases the
recombination rate and lowers the ionisation state of that portion of
gas. The new PN presented in this paper can be seen to satisfy the
first two criteria from the SHS image. The beginning of an asymmetry,
coupled with an increase in brightness, can be clearly seen in
Figure~\ref{hanii}. Our spectroscopic measurements of line ratios (see
Table~\ref{pndata}) further confirm this by showing that the bright
rim does have an enhanced [NII]/H$\alpha$ ratio of 3.2 compared with
the east rim of the nebula where [NII]/H$\alpha$ = 1.5. The ratio for
the whole nebula is [NII]/H$\alpha$ = 1.8. 

{Borkowski} {et~al.} (1990) discuss PN-ISM interactions in terms of
the expansion velocity of the PN and the velocity of the progenitor
star with respect to its surroundings. In a homogeneous, isotropic
ISM, the PN will start to interact on its leading edge. As the PN
expansion causes the nebular density to fall below a critical value,
the gas in the shell is compressed, resulting in a brightness
enhancement in this direction. As the interaction develops the leading
edge of the nebula is slowed by the ISM and the PN shell becomes
distorted as we see in the case of PFP~1.

At first glance this appears to be a reasonable explanation for the
morphology of the new PN, especially as our examination of Galactic
environment found that it sits in a low density region despite its
small z distance of 40 pc. We could therefore conclude that the
interaction is taking place because the PN is moving in that direction
with respect to the local ISM, implying that the CS is moving NW
across the field.

As an alternative cause of the interaction, we also investigated
whether the PN is simply expanding into a region or cloud of enhanced
density to the NW. An IRAS 100$\mu$m dust map does show very faint
dust emission sitting to the south of PFP~1 with a prominence that
curls round the west edge and extends to the interacting rim. The IRAS
data are plotted in Figure~\ref{shassairas} as contours over the
SHASSA H$\alpha$ image. In the absence of distance information for the
emitting dust we cannot tell whether this really is responsible for
the PN interaction or if it is merely a line of sight coincidence.

\section{Summary}

We present preliminary observational findings of an extremely large,
highly coherent, ring-shaped planetary nebula which has just started
to interact with the local ISM. Using a new H$\alpha$ surface
brightness -- radius relation we derive a distance of $\sim$550 pc for
the new PN which is consistent with limits obtained from the inverted
Shlkovsky method. The low reddening, coupled with the lack of features
seen in CO and IRAS dust emission explains why PFP~1 has been able to
expand unencumbered to its present size of $\sim$3.0 pc. We find two
possible causes for the interaction evident on the NW rim: the motion
of the PN with respect to its surroundings as per {Borkowski} {et~al.} 
(1990) or an encounter with a region of enhanced density as it
expands, but are unable to state conclusively which is responsible
based on the available data. An accurate proper motion determination
of the CS candidate would be useful in establishing which of these
scenarios is most likely. Analysis of moderate resolution spectra has
shown that this is a PN of low excitation, which is consistent with
other large-diameter senile examples. A candidate CS has been
identified from UKST B$_J$ and R plates. Follow up spectroscopy is
planned to confirm whether this is the true central star.

The same technique which led to the discovery of PFP~1 has furnished
us with $\sim$30 other candidate PNe of size $>$4\arcmin. A follow up
program is underway to obtain spectra and narrowband imaging of this
sample to be presented in a forthcoming paper (Frew \& Parker, in
preparation).

%
%





\section*{Acknowledgments}

This study used data from the AAO/UKST H$\alpha$ Survey, now available
online as the SuperCOSMOS H$\alpha$ Survey (SHS) by the WFAU of the
Royal Observatory, Edinburgh. These data were supplemented with
calibrated images from the Southern H$\alpha$ Sky Survey Atlas
(SHASSA) and data from the Virginia Tech Spectral-Line Survey
(VTSS) which were produced with support from the National Science
Foundation. This research has made use of the SIMBAD database,
operated at CDS, Strasbourg, France. This publication makes use of
data products from the Two Micron All Sky Survey, which is a joint
project of the University of Massachusetts and the Infrared Processing
and Analysis Center/California Institute of Technology, funded by the
National Aeronautics and Space Administration and the National Science
Foundation. MJP acknowledges support from PPARC and DJF acknowledges
an Australian Postgraduate Award and an ANSTO travel grant which
enabled him to make the observations reported in this paper. QAP
acknowledges support from the AAO and Macquarie University.


\section*{References}



\reference {Acker}, A., {Fresneau}, A., {Pottasch}, S.~R., \& {Jasniewicz}, G. 1998, A\&A,  337, 253

\reference {Aller}, L.~H. 1956, {Gaseous nebulae} (London: Chapman \& Hall, 1956)

\reference {Boffi}, F.~R., \& {Stanghellini}, L. 1994, A\&A, 284, 248

\reference {Borkowski}, K.~J., {Sarazin}, C.~L., \& {Soker}, N. 1990, ApJ, 360, 173

\reference {Brocklehurst}, M. 1971, MNRAS, 153, 471

\reference {Ciardullo}, R., {Bond}, H.~E., {Sipior}, M.~S., {Fullton}, L.~K., {Zhang},  C.-Y., \& {Schaefer}, K.~G. 1999, AJ, 118, 488

\reference {Dame}, T.~M., {Ungerechts}, H., {Cohen}, R.~S., {de Geus}, E.~J., {Grenier},  I.~A., {May}, J., {Murphy}, D.~C., {Nyman}, L.-A., {et al.} 1987,  ApJ, 322, 706

\reference {Dennison}, B., {Simonetti}, J.~H., \& {Topasna}, G.~A. 1998, PASA, 15, 147

\reference {Dopita}, M.~A., \& {Hua}, C.~T. 1997, ApJS, 108, 515

\reference {Dopita}, M.~A., \& {Meatheringham}, S.~J. 1990, ApJ, 357, 140

\reference {Fesen}, R.~A., {Blair}, W.~P., \& {Gull}, T.~R. 1981, ApJ, 245, 131

\reference {Finkbeiner}, D.~P. 2003, ApJS, 146, 407

\reference {Gaustad}, J.~E., {McCullough}, P.~R., {Rosing}, W., \& {Van Buren}, D. 2001,  PASP, 113, 1326

\reference {Haffner}, L.~M., {Reynolds}, R.~J., \& {Tufte}, S.~L. 1998, ApJ, 501, L83

\reference {Hambly}, N.~C., {MacGillivray}, H.~T., {Read}, M.~A., {Tritton}, S.~B.,  {Thomson}, E.~B., {Kelly}, B.~D., {Morgan}, D.~H., {Smith}, R.~E., {et al.} 2001, MNRAS, 326, 1279

\reference {Harris}, H.~C., {Dahn}, C.~C., {Monet}, D.~G., \& {Pier}, J.~R. 1997, in IAU  Symp. 180: Planetary Nebulae, 40

\reference {Hewett}, P.~C., {Irwin}, M.~J., {Skillman}, E.~D., {Foltz}, C.~B., {Willis},  J.~P., {Warren}, S.~J., \& {Walton}, N.~A. 2003, ApJ, 599, L37

\reference {Jacoby}, G.~H. 1981, ApJ, 244, 903

\reference ---. 1989, ApJ, 339, 39

\reference {Jacoby}, G.~H., \& {van de Steene}, G. 1995, AJ, 110, 1285

\reference {Kaler}, J.~B. 1981, ApJ, 250, L31

\reference {Kerber}, F., {Furlan}, E., {Rauch}, T., \& {Roth}, M. 2000, in ASP Conf. Ser.  199: Asymmetrical Planetary Nebulae II: From Origins to Microstructures, 313

\reference {Mallik}, D.~C.~V., \& {Peimbert}, M. 1988, RMxAA, 16, 111

\reference {Melmer}, D., \& {Weinberger}, R. 1990, MNRAS, 243, 236

\reference {Mel'Nik}, A.~M., \& {Efremov}, Y.~N. 1995, AstL, 21, 10

\reference {Mendez}, R.~H., {Kudritzki}, R.~P., {Groth}, H.~G., {Husfeld}, D., \&  {Herrero}, A. 1988, A\&A, 197, L25

\reference {Parker}, Q.~A., {Frew}, D.~J., \& {Stupar}, M. 2004, AAO Newsl., 104, 9

\reference {Parker}, Q.~A., \& {Malin}, D. 1999, PASA, 16, 288

\reference {Parker}, Q.~A., \& {Phillipps}, S. 1998, PASA, 15, 28

\reference {Parker}, Q.~A., \& {Phillipps}, S. 2003, in ASP Conf. Ser. 289: The  Proceedings of the IAU 8th Asian-Pacific Regional Meeting, Volume I, 165--176

\reference {Parker}, Q.~A., {et~al.} 2003, in Proc. IAU Symp. 209, Planetary Nebulae and  their role in the Universe

\reference {Peyaud}, A.~E.~J., {Parker}, Q.~A., \& {Acker}, A. 2003, in SF2A-2003: Semaine  de l'Astrophysique Francaise, 311

\reference {Pottasch}, S.~R. 1996, A\&A, 307, 561

\reference {Rauch}, T. 1999, A\&AS, 135, 487

\reference {Rauch}, T., {Koeppen}, J., \& {Werner}, K. 1994, A\&A, 286, 543

\reference {Savage}, B.~D., \& {Mathis}, J.~S. 1979, Ann.~Rev.~Astron.~Astrophys., 17, 73

\reference {Schlegel}, D.~J., {Finkbeiner}, D.~P., \& {Davis}, M. 1998, ApJ, 500, 525

\reference {Smith}, H. 1976, MNRAS, 175, 419

\reference {Soker}, N., {Borkowski}, K.~J., \& {Sarazin}, C.~L. 1991, AJ, 102, 1381

\reference {Stanghellini}, L., \& {Kaler}, J.~B. 1989, ApJ, 343, 811

\reference {Stanghellini}, L., {Shaw}, R.~A., {Balick}, B., {Mutchler}, M., {Blades},  J.~C., \& {Villaver}, E. 2003, ApJ, 596, 997

\reference {Stone}, R.~P.~S., \& {Baldwin}, J.~A. 1983, MNRAS, 204, 347

\reference {Tweedy}, R.~W., \& {Kwitter}, K.~B. 1996, ApJS, 107, 255

\reference {Tweedy}, R.~W., {Martos}, M.~A., \& {Noriega-Crespo}, A. 1995, ApJ, 447, 257

\reference {Tweedy}, R.~W., \& {Napiwotzki}, R. 1994, AJ, 108, 978

\reference {Wheelock}, S.~L., {Gautier}, T.~N., {Chillemi}, J., {Kester}, D., {McCallon},  H., {Oken}, C., {White}, J., {Gregorich}, D., {et al.} 1994, NASA STI/Recon Tech Report N, 95, 22539

\reference {Xilouris}, K.~M., {Papamastorakis}, J., {Paleologou}, E., \& {Terzian}, Y.  1996, A\&A, 310, 603







\end{document}